\documentclass[aps,groupedaddress]{revtex4}  
\usepackage{graphicx}  
\usepackage{dcolumn}   
\usepackage{bm}        
\usepackage{bbm}
\usepackage{amssymb}   
\usepackage{amscd}   
\usepackage{color}
\usepackage{ulem}
\usepackage{amsmath}
\usepackage{slashbox}

\DeclareMathOperator{\sinc}{sinc}

\begin{document}

\title{A trapped-ion based quantum byte with $10^{-5}$ next-neighbour cross-talk}

\author{Ch.~Piltz}
\author{Th.~Sriarunothai}
\author{A.~ F.~Var\'{o}n}
\author{Ch.~Wunderlich}
\affiliation{Department Physik, Naturwissenschaftlich-Technische Fakult\"at, Universit\"at Siegen, 57068 Siegen, Germany\\ wunderlich@physik.uni-siegen.de}

\date{\today}

\newcommand{\ket}[1]{\left| #1 \right\rangle}               
\newcommand{\bra}[1]{\left\langle #1 \right|}               

\begin{abstract}
The addressing  of a particular qubit within a quantum register is a key prerequisite for scalable quantum computing. In general, executing a quantum gate with a single qubit, or a subset of qubits, affects the quantum states of all other qubits. This reduced fidelity of the whole quantum register could prevent the application of quantum error correction protocols and thus preclude scalability.
We demonstrate addressing of individual qubits within a quantum byte (eight qubits) and measure the error induced in all non-addressed qubits (cross-talk) associated with the application of single-qubit gates. The quantum byte is implemented using  microwave-driven hyperfine qubits of  $^{171}$Yb$^+$ ions confined in a Paul trap augmented with a magnetic gradient field. The measured cross-talk is on the order of $10^{-5}$ and therefore below the threshold commonly agreed sufficient to efficiently realize fault-tolerant quantum computing. Hence, our results demonstrate how this threshold can be overcome with respect to cross-talk.

\end{abstract}

\maketitle

\section*{Introduction}
The precise control of quantum systems is a key ingredient for the realization of quantum information processing devices such as a quantum computer. The level of quantum control achieved with trapped ion based systems is still unsurpassed. Yet, scaling up such a system to enable large-scale quantum computing is a major challenge. If gate errors could be made small enough, then the application of quantum error correction protocols would make scalable fault-tolerant quantum computation possible \cite{Steane1996,Preskill1998}. An important threshold for the tolerable error is $10^{-4}$ per gate \cite{Preskill1998, Knill2010} which has been breached for single-qubit gates with a single trapped ion using microwave radiation \cite{Brown2011,Harty2014}. However, in a register containing several qubits the manipulation of an individual qubit will, in general, induce errors in the quantum state of all  other qubits. This cross-talk may limit the overall fidelity of the quantum register and prevent the application of quantum error correction schemes.

Several methods that allow for addressing of individual ions  have been proposed and demonstrated. By utilizing the micromotion in radiofrequency Paul traps, differential Rabi frequencies between trapped ions were induced  \cite{Turchette1998,Warring2013}. Focused laser beams were used to spatially discriminate ions \cite{Naegerl1999,Schindler2013}, and the use of additional laser beams to reduce cross-talk has been proposed \cite{Shen2013}. 

Static magnetic gradient fields \cite{Mintert2001} that lead to position dependent Zeeman shifts were employed for addressing ions or neutral atoms \cite{Johanning2009,Wang2009,Schrader2004}, and the use of position-dependent light shifts for this purpose was proposed \cite{Staanum2002} and demonstrated \cite{Haljan2005}. Also, differential ac Zeeman shifts \cite{Warring2013} were employed for addressing ions. An inhomogeneous oscillating field that enables spectral resolution of dressed states was demonstrated \cite{Navon2013}.
Also, addressing of neutral atoms confined in an optical lattice was recently demonstrated by the use of additional laser fields and microwave radiation \cite{Lee2013}. 
So far, the cross-talk was always either above the error correction threshold or the system has not yet been proven to be scalable beyond two qubits. 

Here, we present an eight-qubit register - a quantum byte - with next-neighbor cross-talk of the order of $10^{-5}$. It is generally agreed that an error threshold of $10^{-4}$ should be obtained for both single-qubit rotations and multi-qubit gates to efficiently use quantum error correction in a large-scale device. We demonstrate how this threshold can be overcome with respect to cross-talk in an experimental approach that, at the same time, allows for microwave based single-qubit gates with an experimentally demonstrated error per gate of the same order \cite{Brown2011,Harty2014}.
The quantum byte is realized using a string of trapped atomic ions confined in a Paul trap. The qubits are encoded in hyperfine levels of $^{171}$Yb$^+$ ions exposed to a static magnetic field gradient such that each qubit acquires a specific energy splitting between its physical states, thus allowing to distinguish individual qubits by their resonance frequency. We report on precise measurements of cross-talk on all non-addressed ions within the quantum byte using a benchmarking protocol. In this quantum register cross-talk arises mainly from far-detuned electromagnetic pulses that induce spurious rotations in qubits that are not addressed. We also demonstrate, in a three-qubit quantum register, how by optimally choosing addressing frequency and duration of microwave pulses undesired cross-talk could be reduced even further. 
In addition, we briefly discuss the different sources of error that come along with the application of a single pulse and identify non-resonant excitation and microwave light shift as the dominant sources.

\section*{Results}
\subsection*{Addressing a quantum byte}

\begin{figure}[h!]
\includegraphics[width=0.6\columnwidth]{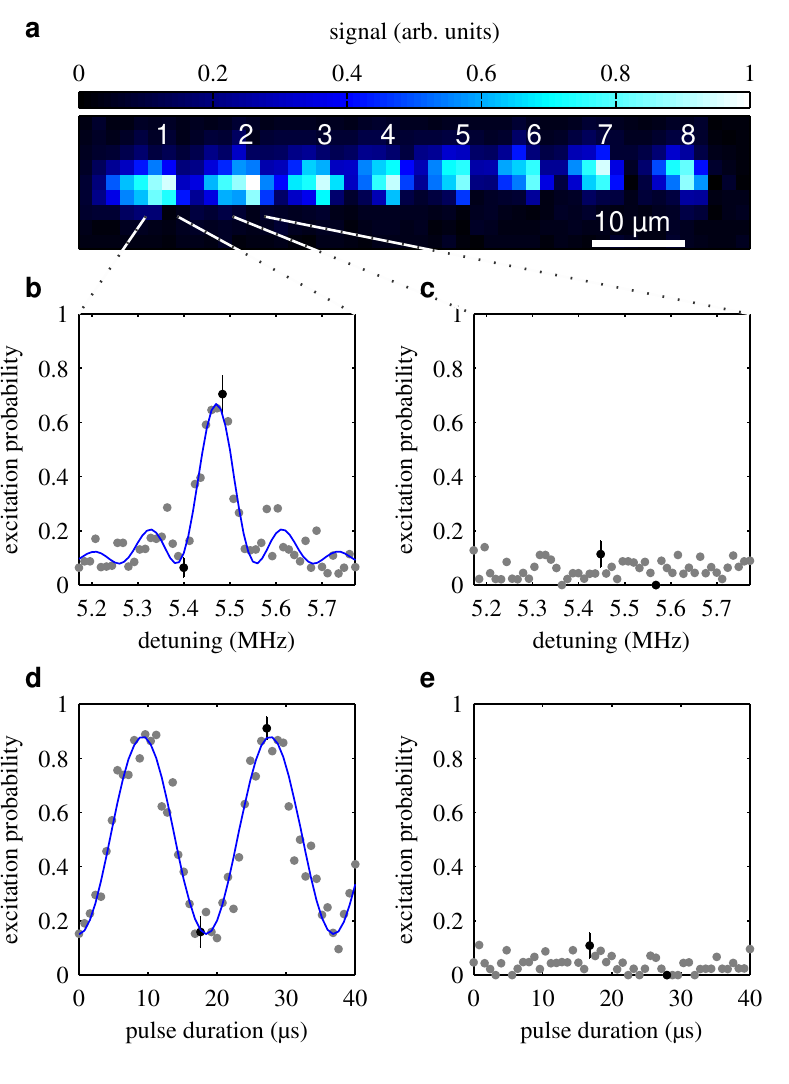}
\caption{Addressing of a single qubit within a quantum byte. (a) Spatially resolved resonance fluorescence (near 369 nm) of eight ions  held in a linear Paul trap detected by an EMCCD camera. 
(b) Microwave-optical double resonance spectrum for a fixed pulse length of 10 $\mu$s serves for determining the  microwave  addressing frequency of an individual ion. Here, the state selective resonance fluorescence signal only in the region of ion 1 is considered. (c) Same as in (b), however measuring the signal in the region of next-neighbor ion 2.
Non-nearest-neighbor ions (3 through 8) are not affected by manipulating qubit 1 either. Their signal is simultaneously measured but not shown for clarity.
 (d) Rabi oscillations are only observed in the region of ion 1 when irradiating all ions at the microwave addressing frequency of ion 1.  (e) Qubit 2 is left virtually unaffected. 
Solid lines represent fits of the data. Two points with error bars are displayed in each graph representing typical statistical standard deviations. Each data point represents 50 repetitions.}
\label{fig:addressing_8qubits}
\end{figure}

We realize a quantum register with a chain of thermally excited $^{171}Yb^+$ ions held in a linear Paul trap. The thermal distribution of vibrational excitation is characterized by the mean vibrational quantum number of the ion crystals axial center-of-mass mode, $ \langle n \rangle \approx 150$ \cite{Khromova2012}. The electronic hyperfine levels of each ion's ground state represent an individual quantum bit:  $\ket{0}\equiv \ket{^2S_{1/2},F=0}$ and $\ket{1}\equiv \ket{^2S_{1/2}, F=1, m_F=+1}$.
The energy of state $\ket{1}$ depends on the magnitude of an external magnetic field (Zeeman effect).  Therefore, a static magnetic gradient field that is created by permanent magnets included in the trap design lifts the degeneracy between the states $\ket{1}$ of different ions. 
The frequency differences $\Delta_{i,j}/2\pi= \nu_i - \nu_j $ between individual ions' $\ket{0} \leftrightarrow \ket{1}$ transition frequencies $\nu_i$ and $\nu_j$ within the chain are given  by $\Delta_{i,j}= g_\mathrm{F} \mu_\mathrm{B} b\, \delta\! z_{i,j} /h$, where $g_\mathrm{F}$ denotes the Land\'{e} $g$-factor, $\mu_\mathrm{B}$ the Bohr magneton, $ b$ the magnitude of the magnetic gradient in the axial trap direction, $\delta\!z_{i,j}$ the spatial separation  between ions $i$ and $j$, and $h$ Planck's constant. The separation  $\delta\! z_{i,j}$ between singly ionized ions is determined by the external axial harmonic trapping potential. In the experiments reported here, this potential  is characterized by a secular  frequency of $2 \pi \times 124$ kHz. As a result, the qubit transition around 12.6 GHz of each ion can be resolved in the frequency domain. A magnetic gradient of 18.8 T m$^{-1}$ leads to differences in the addressing frequencies of next-neighbor qubits of a few MHz (details are given in Methods). 

The ions are initialized in state $\ket{0}$ through optical pumping. Arbitrary single-qubit gates are implemented by the use of microwave pulses near 12.6 GHz. Conditional two-qubit gates between arbitrary qubits within the register can be implemented employing magnetic gradient induced coupling (MAGIC) within the ion chain \cite{Khromova2012,Wunderlich2002,Wunderlich2003}. Read-out of the quantum register is achieved by spatially resolved detection of  resonance fluorescence using an EMCCD detector.

We first determine the resonance frequencies of individual qubits within a register of eight qubits by microwave-optical double resonance spectroscopy [Fig. \ref{fig:addressing_8qubits} (a)]. After preparing the register in state $\ket{00000000}$, microwave pulses of 10 $\mu$s duration and variable frequency 
are applied to all ions. Then, all ions are illuminated with detection laser light and spatially resolved resonance fluorescence is observed. The relative frequency of finding a particular ion in state $\ket{1}$ is obtained upon repeating this sequence of microwave and laser pulses  for a given microwave frequency   [Fig \ref{fig:addressing_8qubits} (b) and (c)]. Thus, observation of individual ion excitation reveals  this qubit's addressing frequency. Here, the chosen pulse duration that does not match the duration for a $\pi$-pulse together with the limited single-shot readout fidelity yield an observed excitation probability below unity in Fig \ref{fig:addressing_8qubits} (b). 

To prove coherent dynamics of a desired single qubit we apply microwave pulses tuned to the respective qubit's addressing frequency while varying the pulse length. We then observe Rabi oscillations of the addressed qubit while the others are virtually left unaffected [Fig. \ref{fig:addressing_8qubits} (d) and (e)] (see Methods for a summary of all ions' addressing and Rabi frequencies.)
The separation between the qubits' addressing frequencies amounts to a few MHz, and is much larger than the Rabi frequency at which an individual qubit is manipulated (typically $2\pi\times 20$ kHz). 
The cross-talk originates, therefore, from the effect of far detuned pulses.

If qubit $i$ is resonantly addressed, the excitation probability of qubit $j$, for $\Omega_j \ll  \Delta_{i,j}$, reads 
\begin{equation}
C_{i,j}=\sin^2(\pi \Delta_{i,j} \tau)(\Omega_j/\Delta_{i,j})^2
\label{eq_Cij_1}
\end{equation}
where $\Omega_j$ is the Rabi frequency of ion $j$ when exposed to the microwave field, and $\tau$ is the duration of the microwave pulse. For typical parameters in our setup the spurious excitation probability is below $10^{-4}$ for next-neighbor qubits and smaller for non-neighboring ions. The fidelity of the final state of qubit $j$ after its spurious evolution due to one pulse applied to qubit $i$ with respect to the initially prepared state $\ket{0}_j$ is given by $1-C_{i,j}$ (Methods).
This work's aim is to measure cross-talk within a quantum register.
We therefore precisely measure spurious excitations by making use of a benchmarking technique that is based on the method of randomized benchmarking \cite{Knill2008,Gambetta2012}. The average state fidelity, $\langle F_{i,j} \rangle$ of qubit $j$ after several sequences of $N$ randomly chosen gates applied to qubit $i$ is given by (Methods)
\begin{equation}
\langle F_{i,j}\rangle(N) = \frac{1}{2} (1 + e^{-2C_{i,j} N}) .
\label{eq_Fidelity}
\end{equation} 
This expression is valid also for the case when a superposition state of qubit $j$ is initially prepared.

\subsection*{Experimentally simulated cross-talk}

\begin{figure}[h!]
\includegraphics[width=0.7\columnwidth]{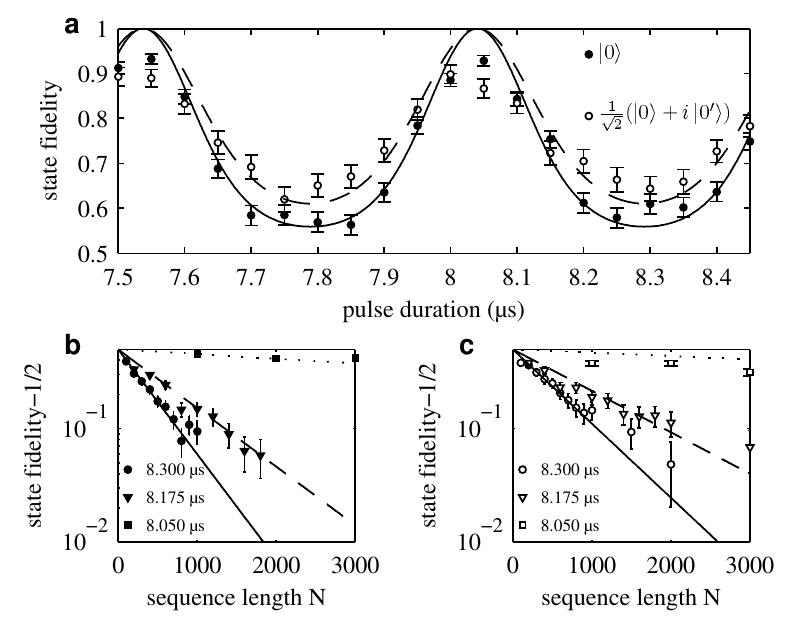}
\caption{Features of cross-talk in single ion measurements. (a) State fidelity after 1000 randomized microwave pulses with variable duration for two different input states ($\ket{0}$: filled cirlces, and $1/\sqrt{2}(\ket{0}+i\ket{0'})$: open circles). 
(b) Fidelity of state $\ket{0}$ as a function of the number of applied pulses. The fidelity of the state decreases exponentially with increasing number of pulses. The decay constant depends on the pulse duration. (c) Similar to (b), but with a superposition state prepared initially.
Lines represent the expectation from the model. Each data point corresponds to 1000 random sequences. Error bars show standard deviations.}
\label{fig:CrosstalkVaryTau1Ion}
\end{figure}

In this section, before turning to the  precise determination of cross-talk in the quantum byte introduced above, we start by  
investigating, with single-ion experiments, the error induced by far detuned microwave pulses. Employing random sequences of gates, we verify the validity of the model that allows to predict the error per single gate as a function of experimental parameters such as the detuning, the power, and the duration of a microwave pulse. 

For this purpose we choose the first order magnetic insensitive levels $\ket{0}$ and $\ket{0'}\equiv\ket{^2S_{1/2},F=1, m_F=0}$ as qubit states. 
Microwave pulses driving this resonance are detuned by $\Delta_\mathrm{\pi}=2\pi\times 2$ MHz from the qubit transition -- a detuning typical for the quantum byte -- and have a variable duration of $7.5 \mu\text{s} \leqslant \tau  \leqslant 8.5$ $\mu$s at a Rabi frequency of $2\pi \times 60.8(5)$ kHz. 
Spurious excitation to the hyperfine states $\ket{^2S_{1/2}, F=1, m_F=\pm 1}$ is negligible here, since a bias field of $0.857 $ mT is applied such that the microwave pulses are detuned by $\Delta_\mathrm{\sigma^+}=-2\pi \times 10$ MHz and $\Delta_\mathrm{\sigma^-}=2\pi \times 14$ MHz from the magnetic $\sigma$ transitions. 

To probe the off-resonant excitation, we apply the following benchmarking protocol. The qubit is initially prepared either in an eigenstate $\ket{\psi}=\ket{0}$, or, by applying a resonant $\pi/2$-pulse, in a superposition state $\ket{\psi}=1/\sqrt{2}(\ket{0}+i\ket{0'})$.
We then apply sequences of up to $N=3000$ detuned pulses where the phase of each pulse is chosen randomly from the set $\{0,\pi/2, \pi, 3\pi/2 \}$.
We vary the pulse duration (that is kept constant during one sequence) and measure the final state of the qubit. When a superposition state was initially prepared, a second resonant $\pi/2$-pulse is applied to the qubit after the random pulse sequence in order to map the superposition state back onto an eigenstate before the projective measurement takes place. The applied protocol, in contrast to the ones used in \cite{Knill2008,Gambetta2012}, does not suffer from single-qubit gate imperfections that otherwise would hinder the detection of small cross-talk errors. In addition, when applied to the quantum byte, it allows to simultaneously measure all cross-talk errors within the register if a certain qubit is addressed.

From the outcome of the projective measurements we deduce the resulting fidelity $F=\bra{\psi}\rho\ket{\psi}$ between the initial state $\ket{\psi}$ and the density matrix $\rho$ after application of a pulse sequence and show the results in figure \ref{fig:CrosstalkVaryTau1Ion} (see Methods for details). 

Fig. \ref{fig:CrosstalkVaryTau1Ion} (a) shows the periodic behavior predicted by equations (\ref{eq_Cij_1}) and (\ref{eq_Fidelity}):  fidelity maxima after 1000 detuned pulses appear at pulse durations $7.55$ $\mu$s and $8.05$ $\mu$s.  
This behavior can be visualized by making use of the Bloch sphere picture. A single detuned microwave pulse rotates the qubit's state vector around an axis close to the z direction in an appropriate rotating reference frame (see Methods for details). The angular velocity of this rotation is given by the generalized Rabi frequency that, for typical experimental parameters, almost equals the detuning. Since the detuning is about 30 times larger than the resonance Rabi frequency, the non-addressed qubit is rotated by angles of $30 \pi$ and $32 \pi$, respectively, for the particular pulse durations mentioned above.
Therefore, after each single pulse, the qubit returns to its initial state (modulo a phase factor $\exp(i\pi)$). In contrast, pulse durations of $7.8$ and $8.3$ $\mu$s rotate the Bloch vector by about $31 \pi$ and $33 \pi$, respectively, which yields the observed fidelity minima.
One may discuss the same situation also in the frequency domain: 
The frequency component of the driving field at the qubits resonance frequency is determined by the Fourier transform of a rectangular pulse (detuned by $\Delta_\mathrm{\pi}$ and with duration of $T$) which is proportional to $\sinc(\Delta_\mathrm{\pi} T/2 \pi)$. For a particular detuning the pulse durations quoted above either lead to a vanishing frequency component that explains the fidelity maxima, or they lead to non-vanishing components that reduce the fidelity.
The effect of detuned pulses also depends on the qubit's state and is smaller for the superposition state [Fig. \ref{fig:CrosstalkVaryTau1Ion}].

In addition, we investigate the rate at which the state fidelity decays during a pulse sequence by varying the sequence length for a given pulse duration.
Figure \ref{fig:CrosstalkVaryTau1Ion} (b) and (c) show the observed decay of the qubit's state fidelity and the prediction from the model (straight lines) for different pulse lengths. Three data sets are shown in each view graph; they are taken for the pulse duration that corresponds to the fidelity maximum, the fidelity minimum, and one value in between, respectively. 

For the pulse duration that corresponds to a rotation of odd multiples of $\pi$, the state fidelity decays most rapidly, while it is better preserved the closer the net rotation is  to an even multiple of $\pi$.
The model (equation (\ref{eq_Fidelity}) predicts an average change of the fidelity per single pulse of $1.06 \times 10^{-3}$ (pulse duration of  8.300 $\mu$s), $5.3 \times 10^{-4}$ (pulse duration of  8.175 $\mu$s), and $1.8 \times 10^{-4}$ (pulse duration of  8.050 $\mu$s)  when state $\ket{0}$ is initially prepared. This is in good agreement with the measured data [Fig. \ref{fig:CrosstalkVaryTau1Ion}].  
For the superposition state [Fig. \ref{fig:CrosstalkVaryTau1Ion} (c)] the fidelity is by a factor of about $\sqrt{2}$ better  than when $\ket{0}$ is prepared initially. The reason is that the benchmarking protocol contains microwave pulses with phases that lead to rotations around the x and y axes. If the qubit is, for example, in an eigenstate of x-rotations, it will be less affected by the detuned pulse. Thus, cross-talk in the quantum register does not only depend on the relative detuning and the Rabi frequencies but also on the register state.

\subsection*{Measured cross-talk within quantum registers}

\begin{figure}[h!]
\includegraphics[width=0.7\columnwidth]{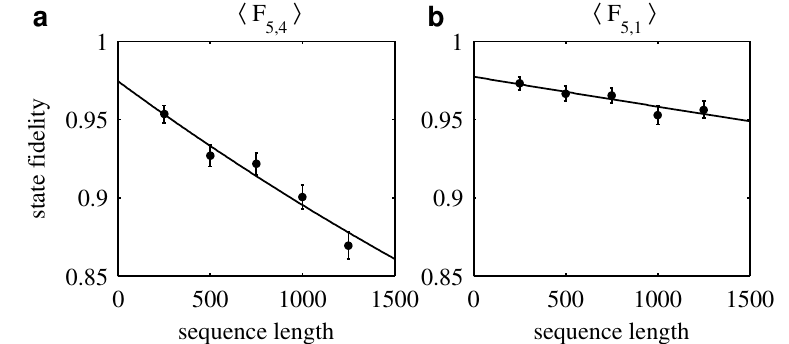}
\caption{Exemplary effect of cross-talk within the quantum byte.
Qubit 5 is addressed and the fidelity decay of the next-neighbor qubit 4 (a) and the non-next-neighbor qubit 1 (b) is observed.
Both qubits' fidelity is initially 0.975(11). 
The dominant next-neighbor cross-talk of $C_{5,4}=7.6(1.3)\times 10^{-5}$ causes qubits 4's fidelity to decay faster than the on of qubit 1 which is affected by a cross-talk of $C_{5,1}=1.9(9)\times 10^{-5}$ only.
Solid lines are a fit to the data from which the cross-talk is deduced. Each point represents 1600 repetitions. Error bars represent standard deviations.}
\label{fig:crosstalk8qubits}
\end{figure}

We now determine the cross-talk within the quantum byte. As before, we apply randomly constructed pulse sequences addressed to one of the qubits within the register and detect the final register state.
Each qubit is encoded in states $\ket{0}$ and $\ket{1}$.
The experimental procedure to deduce the average cross-talk per single gate is a direct extension of the method applied above.
First, the register is initialized in $\ket{00000000}$. Next, a microwave pulse sequence consisting of up to 1250 pulses with randomized phases is applied.  
The frequency and pulse duration of $25 \mu$s are chosen such that they lead to a resonant $\pi$-pulse addressed to one of the qubits at $2\pi \times 20$ kHz Rabi frequency.
At the end of the sequence the register is read out. These experimental steps are repeated while the sequence length is varied.
The cross-talk induced by randomized pulse sequences causes the non-addressed qubits' states to diffuse on the Bloch sphere. A fit of the fidelity, experimentally determined as a function of $N$, gives again the average change of the fidelity per single pulse, the cross-talk $C_{i,j}$ (eq. (\ref{eq_Fidelity})). The model fitted to the data takes into account non-perfect state preparation and detection.

Figure \ref{fig:crosstalk8qubits} shows some exemplary measurement results. The cross-talk of qubit 4 (Fig. \ref{fig:crosstalk8qubits}a)
and 1 (Fig. \ref{fig:crosstalk8qubits}b) are deduced from the data, if qubit 5 is addressed. As one can clearly see the dominant next-neighbor cross-talk causes qubit 4's state fidelity to decay faster than qubit 1's. Recording and analyzing data similar to what is shown in Fig. \ref{fig:crosstalk8qubits} for each pair of qubits allows to construct the cross-talk matrix $\{C_{i,j}\}, i\neq j$. 
We summarize the cross-talk within the quantum byte in table \ref{tab:crosstalk8qubits}. Clearly, next-neighbor cross-talk dominates.
\begin{table}
\caption{Measured cross-talk $\{C_{i,j}\}$ $(\times 10^{-5})$ in the quantum byte.}
\begin{ruledtabular}
\begin{tabular}{l|cccccccc}
\backslashbox{i}{j}& 1& 2 & 3 & 4 & 5 & 6 & 7 & 8\\
\hline
1&- & 3.0(9) & 1.9(8) & 2.2(9) & 2.3(9) & 1.0(8) & 0.6(7) & 0.7(7) \\
2&3.8(1.4) & - & 4.1(1.1) & 2.3(9) & 2.3(1.1) & 1.6(1.1) & 0.9(0.8) & 0.9(0.9) \\
3&2.1(1.0) & 3.7(1.2) & - & 4.5(1.2) & 1.6(7) & 2.1(6) & 0.8(7) & 1.1(6) \\
4&0.9(9) & 1.7(6) & 2.7(1.1) & - & 3.1(9) & 0.8(7) & 0.6(6) & 0.6(6) \\
5&1.9(9) & 1.6(9) & 3.1(1.0) & 7.6(1.3) & - & 3.1(1.0) & 1.8(9) & 0.5(5) \\
6&1.5(5) & 1.2(8) & 1.5(8) & 1.0(8) & 5.5(1.4) & - & 3.6(1.3) & 0.8(8) \\
7&0.8(8) & 1.4(8) & 1.5(7) & 1.2(8) & 1.2(8) & 2.9(1.1) & - & 2.6(8) \\
8&0.8(6) & 1.1(5) & 0.6(6) & 0.8(8) & 2.5(9) & 1.1(8) & 3.4(1.2) & - 
\end{tabular}
\label{tab:crosstalk8qubits}
\end{ruledtabular}
\end{table}

One notices that  the non-next-neighbor cross-talk, if qubit 1 or 2 is addressed is bigger than if, in contrast, qubit 7 or 8 is addressed. The reason for this, at first sight unexpected asymmetry, is the existence of the resonance $\ket{0} \leftrightarrow \ket{0'}$. This resonance is insensitive to the magnetic field to first order, and thus occurs for all ions at about the same frequency. 
An ion in qubit state $\ket{0}$ can, therefore, be spuriously excited to qubit state $\ket{1}$, or to state $\ket{0'}$. Both excitations lead to a reduction of the state fidelity of the qubit $\{\ket{0}, \ket{1}\}$. 
In the experiments reported here, the difference between the addressing frequency of ion 1, $\nu_1$ and the frequency of resonance $\ket{0} \leftrightarrow \ket{0'}$ is 5.5 MHz, determined by a constant bias field of $0.390 $ mT, while $\nu_1-\nu_4=-6.3$ MHz and  $\nu_1-\nu_8=-14.4$ MHz, respectively. Therefore, the excitation on the common resonance ($\ket{0} - \ket{0'}$) is, for this choice of a bias field and the location of the point where the gradient field vanishes, the dominant effect that leads to reduction of the state fidelity. 

Importantly, the experimentally determined cross-talk on the order of $10^{-5}$, as shown in Fig. \ref{fig:crosstalk8qubits} and Table \ref{tab:crosstalk8qubits}, includes {\it any} possible source of cross-talk that affects the state fidelity, with the main contribution arising from non-resonant excitation.

The periodicity of the average cross-talk error per single gate, as shown in figure \ref{fig:CrosstalkVaryTau1Ion} (a), can be exploited to create quantum registers in which the total remaining cross-talk due to non-resonant excitation ideally vanishes or,  at least, is efficiently suppressed. 
The key idea is a  gate applied to one of the qubits that leads to rotations of all non-addressed qubits' Bloch vectors by  about an integer multiple of $2\pi$. Hence, their states will effectively not change (modulo a known phase).
This can be achieved, for example, in a chain of ions in which the detunings of all next-neighbor ions and the Rabi frequencies are equal ($\tilde{\Delta}\equiv \Delta_{i,i+1}$, $i=1, ..., N-1$ and $\Omega_i\equiv \tilde{\Omega}$, $i=1, ..., N$ where $N$ is the number of ions).   A pulse that effectively rotates a non-addressed next-neighbor qubit around $2\pi$ will, to good approximation, also rotate all other non-addressed qubits by the same net rotation as well.\\

\begin{figure}[h!]
\includegraphics[width=0.7\columnwidth]{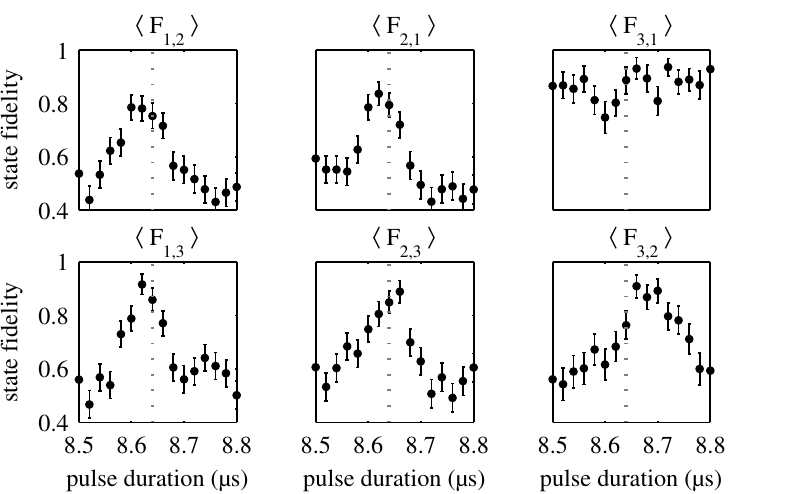}
\caption{Estimating the optimal pulse duration for a three-qubit register. Random sequences of 2000 pulses are addressed to each of the qubits resonantly and the fidelity decay of the other qubits is observed. The optimal pulse duration of $8.64$ $\mu$s (emphasized by a dotted line) yields a preserved state fidelity on all non-addressed qubits. Each data point corresponds to 350 repetitions. Error bars represent one standard deviation.}
\label{fig:figure_threequbits_vary_tau_nm}
\end{figure}

We realize such an optimized quantum register with three ions. 
The constant magnetic gradient yields the same next-neighbor separation in the frequency domain ($\tilde{\Delta}=\Delta_{1,2}=\Delta_{2,3}$). 
A magnetic bias field of about $0.240$ mT results in the same frequency difference between the addressing frequency of ion 1 and the frequency of the resonance $\ket{0} \leftrightarrow \ket{0'}$ ($\tilde{\Delta}=2\pi \times 3.358(3)$ MHz). In this configuration, the magnetic $\sigma^-$ transition of each ion is set apart by an integer multiple of $\tilde{\Delta}$  from the ions' addressing frequencies. 

We first estimate the pulse duration that suppresses the cross-talk, again by application of randomized pulse sequences. The sequence length is kept constant at 2000 pulses while the pulse duration is varied. 
As is evident in figure \ref{fig:figure_threequbits_vary_tau_nm}, a pulse duration of $\tau_{\text{optimal}}=8.64$ $\mu$s results in highly preserved state fidelities for all non-addressed qubits after the application of 2000 pulses.
The difference in the behavior depending on whether qubit 1 or qubit 3 is addressed is, again (as outlined above for the case of the quantum byte), caused by the presence of the resonance $\ket{0} \leftrightarrow \ket{0'}$ that occurs at nearly the same frequency for all three ions. 
The addressing frequency of qubit 1 is closer to this resonance than to the addressing frequency of qubit 3.
This is evident in Fig. \ref{fig:figure_threequbits_vary_tau_nm} where the fidelity decay shows a slower periodicity that is due to a detuning of $\tilde{\Delta}$ only.

The Rabi frequency ($\tilde{\Omega}=2\pi \times 57.9$ kHz) is adjusted such that the pulse duration $\tau_{\text{optimal}}$ results in $\pi$-pulses on the addressed qubit.
According to the experimental procedure described above for the case of a quantum byte, we measure the cross-talk within the optimized three-qubit register by application of up to 5000 pulses. The results are summarized in Table \ref{tab:crosstalk3qubits}.
Again, the resonance $\ket{0} \leftrightarrow \ket{0'}$ is responsible for the increased cross-talk on qubit 2, if qubit 1 is addressed.

\begin{table}
\caption{\label{tab:crosstalk3qubits}Measured cross-talk $C_{i,j}$ $(\times 10^{-5})$ in the optimized three-qubit quantum register.}
\begin{ruledtabular}
\begin{tabular}{l|ccc}
\backslashbox{i}{j}& 1& 2 & 3 \\
\hline
1&- & 23(5) & 6(1) \\
2&8(2) & - & 10(2) \\
3&2.7(3) & 6(1) & -  
\end{tabular}
\end{ruledtabular}
\end{table}

The cross-talk is below the prediction for next-neighbor cross-talk of $3\times 10^{-4}$ according to eq. \ref{eq_Cij_1} at a Rabi frequency of $2\pi \times 57.9$ kHz that is about three times higher than what was used for addressing the quantum byte. This clearly demonstrates the improvement in cross-talk achieved by letting the non-addressed qubits rotate approximately multiples of $2 \pi$. 

Residual cross-talk arises not only because of the possible excitation of the ions' $\ket{0} \leftrightarrow \ket{0'}$ resonance, but also because of addressing frequencies drifting during the experiment \cite{Piltz2013} and through excitation of motional sidebands. Here, the sidebands of the common mode are $\pm 124$ kHz apart from the qubits addressing frequencies \cite{Khromova2012}. The Rabi frequencies on the red and blue sideband transition are proportional to $\sqrt{n}$ and $\sqrt{n+1}$, where $n$ denotes the phonon occupation number of the mode. 
In the  experiments described here, the ions are thermally excited with a phonon occupation of $\langle n \rangle\approx 150$. Since cross-talk is proportional to the square of the transition's Rabi frequency, one can suppress it by further cooling the ions.
The measured values of cross-talk shown in Table \ref{tab:crosstalk3qubits} include all sources of error that affect the state fidelity. 

The method described above can be extended to larger registers in systems with either a spatially varying magnetic gradient, or in systems that allow either anharmonic trap potentials or local potentials as, for example, segmented micro traps \cite{McHugh2005,Lin2009,Hughes2011,Welzel2011,Wilpers2012,Kaufmann2012,Kunert2013}. In these systems the qubits' addressing frequencies and pulse durations can be adjusted appropriately.

\section*{Discussion}
In conclusion, we investigate and characterize cross-talk mechanisms in quantum registers with trapped ions. We experimentally quantify the cross-talk induced by a resonant $\pi$-pulse addressed to a particular qubit by directly measuring the average excitation of all other qubits after up to 1250 pulses.
For the quantum byte, next-neighbor cross-talk is dominant and of order $10^{-5}$. Non-nearest neighbor crosstalk is further suppressed. Importantly, the direct measurement of  cross-talk takes into account all possible sources of error that can affect the state fidelity. Non-resonant excitation was identified as the main source.  

In a three-qubit register a method for further reducing cross-talk is experimentally characterized. This method is based on appropriately adjusting addressing frequencies and pulse durations. Again, cross-talk  is below the error commonly agreed to be sufficient for both single-qubit and multi-qubit gates to efficiently implement fault-tolerant methods for large-scale quantum computing. This demonstrates how we can overcome the threshold with respect to cross-talk even if the single-qubit gates are applied at three times higher Rabi frequencies with the magnetic field gradient remaining constant.

We summarize the different sources of cross-talk and their scaling with experimental parameters in table \ref{tab:estimate}.
\begin{table}
\caption{\label{tab:estimate}Summary of cross-talk mechanisms. Non-resonant excitation,  the microwave light shift, and J-coupling induce cross-talk during single-qubit operations. Microwave light shift and J-coupling affect superposition states only. The light shift can be accounted for, and was compensated in the measurements reported here. The value for non-resonant excitation, $C$ quoted in this table is the average next-neighbor cross-talk within the quantum byte (compare Tab. \ref{tab:crosstalk8qubits}).  The error due to microwave light shift is measured for a Rabi frequency of $2\pi \times 20$ kHz and detuning of $2 \pi \times 2$ MHz which are typical parameters of the quantum-byte. The error due to J-coupling is measured for the case of a three-qubit register  ($b=18.8$ Tm$^{-1}$ and $\nu_1=2\pi \times 123.5(2)$ kHz) and can be considered as an upper bound. 
The rightmost column shows the scaling of a given error when rectangular pulses with Rabi frequency $\Omega$ are applied to ions confined in a harmonic trapping potential with secular frequency $\nu_1$ and a magnetic gradient $b$.}
\begin{ruledtabular}
\begin{tabular}{lll}
source of cross-talk & measured value & scaling with experimental parameters\\
\hline
non-resonant excitation & $C=3.8(1.3) \times 10^{-5}$ & $\Omega^2 b^{-2}\nu_1^{4/3}$\\
microwave light shift   & $\epsilon_\mathrm{ac}=5.9(7) \times 10^{-5}$ & $\Omega^2 b^{-2} \nu_1^{4/3}$ \\
J-coupling              & $\epsilon_\mathrm{J}=4.3(8) \times 10^{-6}$ & $\Omega^{-2} b^4 \nu_1^{-4}$ 
\end{tabular}
\end{ruledtabular}
\end{table}

The state fidelity of superposition states is, in addition to non-resonant excitation, affected by the light shift induced by a non-resonant field and J-coupling among individual ions. Both of these effects change the phase of a superposition state by $\delta \varphi_\mathrm{ac,J}$ during the pulse duration $\tau$.  The phase shift will reduce the state fidelity to
$F=\frac{1}{2}(1+\cos(\delta \varphi_\mathrm{ac,J}))$ and therefore the induced error is $\epsilon_\mathrm{ac,J}=1-F \approx (\delta \varphi_\mathrm{ac,J}/2)^2$. 

In case of the light shift the phase change is $\delta \varphi_\mathrm{ac}=\frac{\Omega^2}{2 \Delta} \tau$. The frequency difference $\Delta=b \delta z$ depends on the magnetic gradient $b$ and the spatial separation of ions, $\delta z$ which depends in a harmonic trapping potential on the secular frequency $\nu_1$ as $\delta z \propto \nu_1^{-2/3}$ \cite{James1998}. The pulse duration is inversely proportional to the Rabi frequency and hence the error induced by light shift scales as $\epsilon_\mathrm{ac}\propto \Omega^2 b^{-2} \nu_1^{4/3}$. We precisely deduce the size of the light shift by using the single-ion quantum lock-in amplifier technique \cite{Kotler2011}. For parameters that are typical for the investigated registers (Rabi frequency of $2\pi \times 20$ kHz and detuning of $2\pi \times 2$ MHz),  the light shift is measured as $2\pi \times 98(6)$ Hz. During a resonant $\pi$-pulse this would induce a state infidelity of $5.9(7) \times 10^{-5}$. This effect is systematic and was suppressed in the study by the use of a spin-echo pulse (Methods). During the execution of a quantum algorithm, one can account for it by adjusting the phase of subsequent pulses \cite{Warring2013}.

J-coupling among individual ions,too, changes the phase of a superposition state. The induced phase shift is proportional to the coupling constant $J$ with $J \propto b^2 \nu_1^{-2}$ and hence the induced error scales as $\epsilon_\mathrm{J}\propto b^4 \nu_1^{-4} \Omega^{-2}$. For the case of the three-qubit register the measured next-neighbor coupling strength in our experimental setup is $2\pi \times 33(3)$ Hz \cite{Khromova2012} which yields, if not compensated for, an induced error of $4.3(8) \times 10^{-6}$. 

We briefly discuss the scaling of the different sources of cross-talk.  
Cross-talk due to non-resonant excitation, $C\propto \Omega^2 b^{-2}\nu_1^{4/3}$ ( eq. (\ref{eq_Cij_1})) could be further suppressed by increasing the magnetic gradient and/or by reducing the power of the microwave pulses that implement single-qubit gates. 
At the same time a reduced microwave power would also reduce the systematic effect superposition states experience due to microwave light shift.
A reduction of the external axial trapping potential would also reduce the cross-talk errors of these two sources but the errors scale only as $\nu_1^{4/3}$. 
The error due to undesired J-coupling would, in contrast, increase if the resonant Rabi frequency is reduced because the effect would longer act on the qubits during the longer gate duration. 

A different approach to reduce cross-talk due to non-resonant excitation would be the use of more elaborate pulse shapes \cite{Timoney2008}. For example, a Gaussian pulse covers a narrower span in frequency, as compared to a square pulse,  and therefore reduces off-resonant excitation even further.

Another possibility to elude the spurious effect of cross-talk is to calibrate it precisely. Once the spurious systematic rotation that is induced by a single pulse is known, one can either compensate it or take it explicitly into account \cite{Aude2013}. For example, the systematic rotations around the qubit's $z$ axis due to microwave light shift can be compensated by adjusting the relative phase of consecutive microwave pulses and do not require the application of additional pulses.

Spurious excitation on undesired hyperfine transitions can be further suppressed by either increasing the magnetic bias field or by adjusting the microwave polarization.
For $\pi/2$-rotations around the $\pm x$ and $\pm y$ axes that together with  $\pi$-pulses generate the single-qubit Clifford group, the cross-talk mechanisms are the same as for $\pi$-pulses investigated in this study. The pulse duration is halved but the cross-talk is still of the same order of magnitude.

Using magnetic gradient induced coupling to carry out an entangling two-qubit gate is realized by application of two $\pi/2$-pulses and a conditional evolution time \cite{Khromova2012,Piltz2013}. Hence, a non-addressed qubit would suffer from the application of two single-qubit gates only. The cross-talk regarding this gate would therefore be of the same order of magnitude as the single-qubit gates studied here in detail.

\section*{Methods}

\subsection*{Cross-talk for far detuned pulses}
The cross-talk in the investigated addressing scheme is mainly due to the effect of far detuned pulses of coherent electromagnetic radiation. While one qubit $i$ is resonantly addressed at microwave frequency $\omega_i$, the other qubits $j\neq i$ are exposed to non-resonant radiation at frequency $\omega_i
 + \Delta_{i,j}$. We are interested in describing the spurious dynamics of a given qubit $j$ while qubit $i$ is addressed. 

First we concisely review relevant features of the quantum dynamics of a single qubit under irradiation with microwave radiation. The time evolution of qubit $j$ is conveniently described in a  reference frame that rotates at frequency  $\omega_i$. 
Specifically, the evolution of qubit $j$ while irradiated by microwave radiation is described by a rotation 
$R_\mathrm{\bf n}(\theta)=\exp(-i\theta/2\mathrm{\bf n}\cdot\mathrm{\boldsymbol \sigma}) \in \mbox{SU(2)}$ around unit vector $\mathrm{\bf n}$ by an angle $\theta=\Omega_\mathrm{R} \tau$, where $\Omega_\mathrm{R} = \sqrt{\Omega_i^2+\Delta_{i,j}^2}$ denotes the generalized Rabi frequency, $\Delta_{i,j}$ is the detuning, $\tau$ is the duration of a square pulse, and $\mathrm{\boldsymbol \sigma} =(\sigma_\mathrm{x}, \sigma_\mathrm{y},\sigma_\mathrm{z})$ with the Pauli matrices $\sigma_i$. 
The magnitude of the component of $\mathrm{\bf n}$ in the $xy$-plane of the Bloch sphere, $n_\perp= \Omega_i / \Omega_\mathrm{R}$, and in the z-direction, $n_\mathrm{z}= \Delta_{i,j} / \Omega_\mathrm{R}$. Thus,  for far detuned pulses, the qubit's dynamics on the Bloch sphere while exposed to a far detuned pulse  is described by a rotation around an axis that is  nearly parallel to the z axis.  

In a quantum register where all qubits have been prepared in their ground state $\ket{0}$ the addressing of a single qubit $i$ for the duration $\tau$ will spuriously excite all other qubits $j\neq i$. The induced cross-talk error per single pulse on qubit $j$ is 
\begin{eqnarray}
C_{i,j}&=&| \bra{1}  R_\mathrm{\bf n}(\theta) \ket{0} |^2\\ \nonumber
           &=& (\Omega_i/\Omega_\mathrm{R})^2 \sin^2(\Omega_\mathrm{R} \tau/2).
\label{eq:crosstalkerror}
\end{eqnarray}
For far detuned pulses ($\Delta \gg \Omega$) 
\begin{eqnarray}
C_{i,j} \approx (\Omega_i/\Delta_{i,j} )^2 \sin(\Delta_{i,j} \tau).
\label{eq:crosstalkerror_app}
\end{eqnarray}
Thus, the maximal spurious excitation probability for a single pulse is $(\Omega_i/\Delta_{i,j} )^2$. 
For a pulse duration of $\tau=2\pi/\Delta_{i,j}$  the qubit's Bloch vector is rotated by an angle of $2\pi$ and hence leaves the qubit excitation probability effectively unchanged. One can take advantage of this periodicity to further lower the cross talk error (compare Figs. \ref{fig:figure_threequbits_vary_tau_nm}).

When a superposition state of qubit $j$ is prepared initially (Fig. \ref{fig:CrosstalkVaryTau1Ion} (a) and (c)), the effect of detuned pulses on this qubit is probed employing a Ramsey-type experiment. After preparing the qubit in $\ket{0}$ a resonant $\pi/2$-pulse is applied on the $\ket{0} - \ket{0'}$ resonance to drive qubit  $j$ into a superposition state before qubit $i$ is addressed. The second $\pi/2$-pulse would ideally excite qubit state into state $\ket{1}$, if qubit $j$ were not affected by the non-resonant pulses addressing qubit $i$. The deviation from full population transfer is given by equation (\ref{eq:crosstalkerror_app}). 

\subsection*{Estimating state fidelities}

We experimentally deduce cross-talk in the quantum register by observing the state fidelity of the non-addressed qubits $j \neq i$. This fidelity decays during the application of the benchmarking protocol addressed to qubit $i$ (eq. (\ref{eq:fidelitydecay})).
The fidelity 
\begin{eqnarray}
	\langle F_{i,j} \rangle=\bra{\psi_j}\rho_{i,j}\ket{\psi_j}.
	\label{eq:fidelity}
\end{eqnarray}
Here, $\rho_{i,j}$ denotes the final state density matrix of qubit $j$ after having been exposed to random sequences of detuned pulses addressed to qubit $i$, and $\ket{\psi_j}$ is the initial state of qubit $j$.

With respect to the initially prepared ground state, $\ket{0}$, the fidelity of the final state is 
\begin{equation}
	\langle F_{i,j} \rangle=\bra{0}\rho_{i,j}\ket{0}=\rho_{i,j}^{00}.
\end{equation}
We can, therefore, determine the average fidelity of qubit $j$ by measuring the excitation probability (or rather relative frequencies) of qubit $j$ into $\ket{1}$, $P_{i,j}$, after the benchmarking sequence has been addressed to qubit $i$, and
\begin{eqnarray}
	\langle F_{i,j}\rangle=1-P_{i,j}.
\end{eqnarray}

With respect to the superposition state $\frac{1}{\sqrt{2}}(\ket{0}+e^{i\pi/2}\ket{1})$
the fidelity (\ref{eq:fidelity}) reads
\begin{eqnarray}
\langle F_{i,j} \rangle&=&\frac{1}{2}(1+2|\rho_{i,j}^{01}|).
\label{eq:fidelitysuperposition}
\end{eqnarray}
The probability $P_{i,j}$ to find qubit $j$ in state $\ket{1}$ after the application of the Ramsey-type pulse sequence mentioned above (the relative phase between the $\pi/2-$ pulses is not varied) is  
\begin{equation}
P_{i,j}=\frac{1}{2}(1+2|\rho_j^{01}|) = \langle F_{i,j} \rangle.
\end{equation}

\subsection*{Effect of randomized pulse sequences}

\begin{figure}[h!]
\includegraphics[width=0.6\columnwidth]{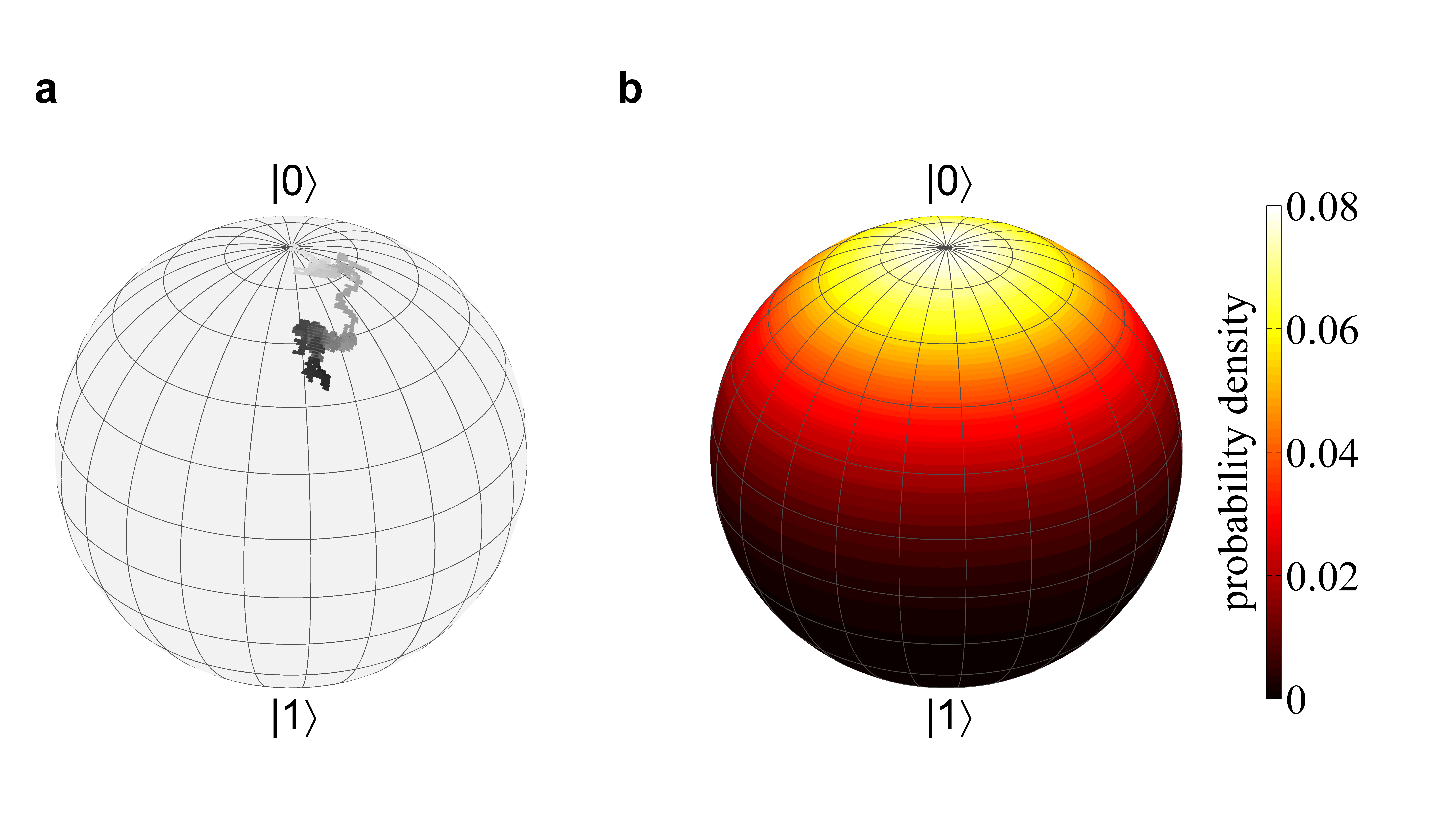}
\caption{Effect of randomized pulse sequences. (a) A randomized sequence of $N=1250$ pulses with a Rabi frequency of $2\pi \times 20$ kHz and a detuning of $2\pi \times 2$ MHz causes the state vector which is initialized in $\ket{0}$ to perform a random walk on the Bloch sphere. After each pulse the state is indicated by a dot where the shading indicates the pulse number (from light at the beginning to dark at the end). (b) For several benchmarking sequences (with same parameters as in (a)) the average final state is distributed on the Bloch sphere. This can be treated as a diffusion problem and the resulting probability density is shown.}
\label{fig:randomwalk_diffusion}
\end{figure}

For typical parameters used in these experiments ($\Omega=2\pi\times 20$ kHz and $\Delta_{i,i+1}=2\pi\times 2$ MHz), the cross-talk  (\ref{eq:crosstalkerror}) is of order $10^{-5}$ which is much smaller than typical state preparation and detection errors. To measure such a small effect we make use of randomized gate sequences \cite{Knill2008}. From the application of several gates and the observation of accumulating errors that yield a decay of the state fidelities we deduce the error per single gate. In what follows we describe the decay of state fidelities due to far detuned randomized pulse sequences. We derive an analytical formula that  describes the average fidelity after applying sequences of a given length consisting of  pulses with given detuning and microwave power.

According to the discussion above a single microwave pulse, which realizes a gate addressed to qubit $i$, will rotate the state vectors of the qubit $j$ about an axis defined by the phase of the pulse and the mutual detuning $\Delta_{i,j}$.
A sequence of pulses with randomly chosen phases will, therefore, result in a random walk of qubit $j$ on its Bloch sphere [Fig. \ref{fig:randomwalk_diffusion} (a)].

Since the spin changes its state only slightly during one pulse 
applied at time $t_n$, $n=1,2,3,...N$ that is part of a benchmarking sequence of $N\gg 1$ gates, we treat the process as a continuous motion. If the random paths of many sequences are taken into account, the average position of the final state of qubit $j$ will be spread over the Bloch sphere [Fig. \ref{fig:randomwalk_diffusion} (b)] which can be described, in general, by the diffusion equation
\begin{eqnarray}
\partial_t \phi_{i,j}(\mathrm{\bf r},t)=D_{i,j}\nabla^2 \phi_{i,j}(\mathrm{\bf r},t).
\label{eq:diffusion}
\end{eqnarray}
Here, $\phi_{i,j}(\mathrm{\bf r},t)$ denotes the probability density function of finding qubit $j$'s state vector at the location $\mathrm{\bf r}$ on its Bloch sphere at time $t$, and $D_{i,j}$ denotes the diffusion constant.

Since we are interested in describing the state fidelities rather than the exact position on the Bloch sphere, we simplify the diffusion problem to a single spatial coordinate $x$ on the Bloch sphere. This coordinate is the angle between the initial state vector and the state vector after a benchmarking sequence has been applied. In case of the energy eigenstate $\ket{0}$ beeing the initial state, for example, this coordinate would be the latitude of the Bloch sphere. The reason is that the fidelity between the energy eigenstate and all states on the same latitude is the same. We solve the simplified problem for the state vector being either an energy eigenstate or a superposition state. For both cases a corresponding polar coordinate system is defined such that the initial state is found at $x_0=\pi$.
In addition, these coordinate systems feature periodic boundary conditions $\phi(x+2\pi)=\phi(x)$ because of the Bloch sphere's periodicity. 
In these coordinate systems the fidelity between the states at $x$ and the initial state is described by $F(x)=\frac{1}{2}(1-\cos(x))$.

The 1-dimensional  diffusion equation is solved for both cases by
\begin{eqnarray}
\phi_{i,j}(x,t)=\frac{1}{\pi}\sum_{m=0}^{\infty}e^{-D_{i,j}t/m^{2}}(-1)^m\cos(mx)-\frac{1}{2\pi}.
\end{eqnarray}
In fig. \ref{fig:randomwalk_diffusion} (b) a stationary solution for the state initialized in $\ket{0}$ is illustrated.
With knowledge of the probability densitiy, one can calculate the time-dependent state fidelity with the initial state:
\begin{eqnarray}
	\langle F_{i,j}(t)\rangle &=& \int_0^{2\pi}dx \phi_{i,j}(x,t) F(x),\nonumber\\
	&=& \frac{1}{2} (1 + e^{-D_{i,j}t}).
\end{eqnarray}

For a one dimensional random walk of $N$ steps with a hopping distance $\delta x_{i,j}$ during a single time step $\tau$ the diffusion constant is $D_{i,j}=\frac{1}{2}\frac{{\delta x}_{i,j}^2}{\tau}$ (e. g. \cite{Mahnke2009}). The total time of the random walk is $t=\tau N$ and each step causes a fidelity change $C_{i,j} = 1 - F(\delta x) \approx \frac{{\delta x}_{i,j}^2}{4}$. 
Hence, we conclude $D_{i,j} t = 2 C_{i,j} N$.

In summary, the state diffusion of qubit $j$ is driven by cross-talk induced by pulse sequences addressed to qubit $i$. The average fidelity of qubit $j$ after qubit $i$ was addressed with $N$ random pulses is related to the error probability per single gate $C_{i,j}$ by 
\begin{eqnarray}
	\langle F_{i,j}(N)\rangle &=& \frac{1}{2} (1 + e^{-2C_{i,j} N}).
	\label{eq:fidelitydecay}
\end{eqnarray}
 
Therefore, the cross-talk $C_{i,j}$ per single gate can be deduced from measuring the fidelity $\langle F_{i,j}(N)\rangle$ of a non-addressed qubit $j$ while the benchmarking protocol is applied to qubit $i$.

However, a limited single-shot readout fidelity and slightly imperfect initial state preparation causes the detected state fidelity to deviate from unity even in the absence of benchmarking gates (N=0).
In the model fitted to the data
\begin{eqnarray}
f(p_0,p_1,N) = 1/2\times(1+(2 p_0-1)\exp(-2/p_1\times N))
\end{eqnarray}
the parameter $p_0<1$ takes these imperfections into account. The free parameter $p_1=2/C_{i,j}$ describes the decay of fidelity (from which the cross-talk is deduced) and $N$ is the number of  pulses applied to qubit $i$.

\subsection*{Experimental procedures}
We briefly describe the procedure that was carried out to experimentally deduce the cross-talk  $C_{i,j}$.
The results for the input state $\ket{0}$ (Fig. \ref{fig:CrosstalkVaryTau1Ion} (a) and (b), Fig. \ref{fig:crosstalk8qubits} (a) and (b) and Fig. \ref{fig:figure_threequbits_vary_tau_nm}) are obtained by the following experimental procedure. 
The ions are initialized in state $\ket{0}\equiv \ket{^2S_{1/2},F=0}$ through optical pumping on the optical transition $\ket{^2S_{1/2}, F=1} \leftrightarrow \ket{^2P_{1/2},F=1}$ using laser light near 369 nm.
Then, a randomized microwave pulse sequence (described above) is applied.
In the applied benchmarking protocol the phase and therefore the nutation axis of each pulse is chosen randomly from $\{0,\pi/2,\pi,3\pi/2\}$ in every single realization of the sequences. All the pulses that are applied are of rectangular shape and the duty cycle of each sequence is $50\%$.
Doppler cooling of the ion chain and state selective detection of the ions' qubit state is achieved by driving the transition $\ket{^2S_{1/2},F=1} \leftrightarrow \ket{^2P_{1/2},F=0}$ again with laser light near 369 nm. Optical pumping into the metastable D$_{3/2}$ state is prevented by illuminating the ions with laser light near 935 nm \cite{Balzer2006}. These two light fields are referred to as detection laser light in this paper. 
Read-out of the quantum register is achieved by spatially resolved detection of  resonance fluorescence using  an EMCCD detector. Alternatively, a photomultiplier may be used to detect the overall fluorescence. 
These steps are repeated in order to obtain statistical significance. 
From the probability of finding an ion in its bright state the fidelity is deduced (see above).

The results with the input state being a superposition state (Fig. \ref{fig:CrosstalkVaryTau1Ion} (a) and (c)) are optained by a Ramsey-type experiment. The qubit is prepared in the superposition state by a resonant $\pi/2$-pulse on the $\ket{0}\equiv \ket{^2S_{1/2},F=0} \leftrightarrow \ket{0'}\equiv \ket{^2S_{1/2},F=1,m_F=0}$ transition before a benchmarking sequence is applied.
Just before the detection a second Ramsey $\pi/2$-pulse on the $\ket{0} - \ket{0'}$ resonance is applied to probe the state.
In order to suppress any effect of a possible precession that originates from an imperfect preparation of the superposition state due to a slightly detuned $\pi/2$-pulse, a resonant spin echo $\pi$-pulse on the $\ket{0} - \ket{0'}$ resonance is applied between the two resonant $\pi/2$-pulses. 
Since this spin echo pulse also compensates the microwave light shift its effect was independently measured as discussed above.

\subsection*{Single-qubit gates}
Using our benchmarking protocol we derive a lower bound for the error of each addressed single-qubit gate of $5 \times 10^{-3}$. In future work microwave based high-fidelity single-qubit gates \cite{Brown2011,Harty2014} will be implemented by making necessary technical improvements while keeping cross-talk still below the error limit that is associated with fault-tolerant quantum computing.
On the way to such a realization there may be a trade-off between the methods that reduce cross-talk and the methods that improve single-qubit gate fidelities as discussed in the main text. 

Table \ref{tab:addressing} lists the experimentally determined qubit resonance- and Rabi frequencies for the quantum byte used here. 

\begin{table}
\caption{\label{tab:addressing}Addressing of the quantum byte. These results are obtained employing microwave-optical double resonance spectroscopy (Fig. \ref{fig:addressing_8qubits}). The variation in Rabi frequencies in the eight-ion experiment reported here is due to an inhomogeneous static magnetic field perpendicular to the trap axis. Ion $i=1,2,3, ..., 8$ was addressed using a Rabi frequency $\Omega_i= 2\pi\times 20$ kHz.  The perpendicular magnetic field component was nulled for the experiments demonstrating the optimized three-qubit register.}
\begin{ruledtabular}
\begin{tabular}{ccc}
ion number & resonance frequency & Rabi frequency  \\
 $i$& $\nu_i $ (GHz)& $\Omega / 2\pi$ (kHz)\\
\hline
1 & 12.648298(3) & 54.1(5)\\
2 & 12.650634(3) & 46.7(6)\\
3 & 12.652658(4) & 43.5(6)\\
4 & 12.654523(4) & 40.0(7)\\
5 & 12.656375(9) & 38.2(1.2)\\
6 & 12.658282(9) & 34.4(6)\\
7 & 12.660302(9) & 30.7(8)\\
8 & 12.662694(11) & 28.5(8)\\
\end{tabular}
\end{ruledtabular}
\end{table}

\section*{Acknowledgement}
We acknowledge funding by the Bundesministerium f\"ur Bildung und
Forschung (FK 01BQ1012), from the European Community's Seventh
Framework Programme (FP7/2007-2013) under Grant Agreement No. 270843
(iQIT), and from Deutsche Forschungsgemeinschaft.

\section*{Author contributions}
Ch. W. proposed the project and conceived the idea. A. F. V. and Ch. P. designed the experiment and carried out the theoretical study. Th. S. and Ch. P. carried out the experimental study. Ch. P. and Ch. W. wrote the paper. All authors analysed the data and commented on the manuscript.

\end{document}